\documentclass[a4paper,10pt,twoside]{cpc-hepnp}

\usepackage{multicol}
\usepackage{multirow}
\usepackage{graphicx}
\usepackage{booktabs}
\usepackage{amssymb,bm,mathrsfs,bbm,amscd}
\usepackage[tbtags]{amsmath}
\usepackage{lastpage}

\begin{document}

\fancyhead[c]{\small Chinese Physics C~~~Vol. XX, No. X (201X)
XXXXXX} \fancyfoot[C]{\small 010201-\thepage}

\footnotetext[0]{Received 14 March 2009}

\title{Preliminary Research on Dual-Energy X-Ray Phase-Contrast Imaging\thanks{Supported by the Major State Basic Research Development Program (2012CB825800), the Science Fund for Creative Research Groups (11321503) and the National Natural Science Foundation of China (11179004, 10979055, 11205189, 11205157). }}

\author{%
      HAN Hua-Jie(韩华杰)$^{1,2}$
\quad WANG Sheng-Hao(王圣浩)$^{1}$
\quad GAO Kun(高昆)$^{1;1)}$\email{gaokun@ustc.edu.cn}%
\quad WANG Zhi-Li(王志立)$^{1}$\\
\quad ZHANG Can(张灿)$^{1}$
\quad YANG Meng(杨萌)$^{1}$
\quad ZHANG Kai(张凯)$^{3;2)}$\email{zhangk@ihep.ac.cn}%
\quad ZHU Pei-Ping(朱佩平)$^{3}$
}
\maketitle

\address{%
$^1$ National Synchrotron Radiation Laboratory, University of Science and Technology of China, Hefei 230027, China\\
$^2$ School of Engineering Science, University of Science and Technology of China, Hefei 230027, China\\
$^3$ Institute of High Energy Physics, Chinese Academy of Sciences, Beijing 100049, China\\
}

\begin{abstract}
Dual-energy X-ray absorptiometry (DEXA) has been widely applied to measure bone mineral density (BMD) and soft-tissue composition of human body. However, the use of DEXA is greatly limited for low-Z materials such as soft tissues due to their weak absorption. While X-ray phase-contrast imaging (XPCI) shows significantly improved contrast in comparison with the conventional standard absorption-based X-ray imaging for soft tissues. In this paper, we propose a novel X-ray phase-contrast method to measure the area density of low-Z materials, including a single-energy method and a dual-energy method. The single-energy method is for the area density calculation of one low-Z material, while the dual-energy method is aiming to calculate the area densities of two low-Z materials simultaneously. Comparing the experimental and simulation results with the theoretic ones, the new method proves to have the potential to replace DEXA in area density measurement. The new method sets the prerequisites for future precise and low-dose area density calculation method of low-Z materials.
\end{abstract}



\footnotetext[0]{\hspace*{-3mm}\raisebox{0.3ex}{$\scriptstyle\copyright$}2013
Chinese Physical Society and the Institute of High Energy Physics
of the Chinese Academy of Sciences and the Institute
of Modern Physics of the Chinese Academy of Sciences and IOP Publishing Ltd}%

\begin{multicols}{2}

\section{Introduction}

The idea of dual-energy X-ray absorptiometry can be traced back to the 1970s, when the method of differentiating low-Z materials from high-Z materials with a dual-energy scan was firstly proposed by Hounsfield \cite{lab1}. Following his work, researchers including Zatz, Alvarez et al. \cite{lab2, lab3, lab4} reported the advantages of dual-energy method in material differentiation. In 1976, R. A. Rutherford utilized dual-energy scan to determine the effective atomic number and electron density of materials, and applied it to human tissue imaging \cite{lab5}. Recently, dual-energy absorptiometry has found increasingly wide applications in the fields of clinical diagnosis, public security, etc \cite{lab6, lab7, lab8, lab9, lab10, lab11, lab12}. As an application, dual-energy X-ray absorptiometry is the “golden standard” in bone mineral density (BMD) measurement, whereby percent fat in soft tissue, fat mass, lean tissue mass can also be obtained \cite{lab6, lab7, lab8, lab9}.

However, low-Z materials, such as soft tissues, show weak absorption of X-ray. Besides, different soft tissues show similar characters in X-ray absorption \cite{lab13}. Therefore conventional absorption-based dual-energy CT has an inherent defect in material differentiation for weak absorption materials \cite{lab14}. While X-ray phase-contrast imaging offers a dramatically higher contrast for weak absorbers as reported in the past few decades \cite{lab15, lab16,lab17,lab18,lab19,lab20,lab21}, the reason lies in that the refraction coefficient exceeds the absorption coefficient with at least three orders of magnitude. Moreover, the refraction coefficients of different soft tissues vary greatly \cite{lab15}.

In this paper, an innovative dual-energy phase-contrast method is proposed, which is able to remarkably promote the quality of identification and differentiation of weakly absorbing materials, and meanwhile decrease the radiation dose. Firstly, we derive the equations for area density calculation, including single-energy method and dual-energy method. Secondly, based on the experimental results and a numerical simulation, we demonstrate the feasibility of the methods. Finally, we discuss experimental data and potential application of the new dual-energy phase-contrast imaging method.

\section{Theories}

When a beam of X-ray penetrates an object, the variation of the intensity is given by
\begin{eqnarray}
\label{eq2}
I = {I_0}{e^{ - \mu t}}.
\end{eqnarray}
where ${I_0}$ is the intensity of the incident X-ray, $I$ represents the intensity of the emergent X-ray, $t$ is the transmission length, and $\mu $ is the linear attenuation coefficient. Generally, Eq. (1) is written as
\begin{eqnarray}
\label{eq2}
I = {I_0}{e^{ - (\mu /\rho )\rho t}}.
\end{eqnarray}
where $\rho $ is the density of the substance, ${\mu  \mathord{\left/
 {\vphantom {\mu  \rho }} \right.
 \kern-\nulldelimiterspace} \rho }$ is known as mass attenuation coefficient, which is merely determined by the characters of the substance. And $\rho t$ is the area density of substance, denoted by $M$. The area density thus is given by
 \begin{eqnarray}
\label{eq2}
M =  - \frac{{\ln (I/{I_0})}}{{{\mu  \mathord{\left/
 {\vphantom {\mu  \rho }} \right.
 \kern-\nulldelimiterspace} \rho }}}.
\end{eqnarray}

The precise measurement of BMD is crucial for the diagnosis of osteoporosis \cite{lab22}, and dual-energy X-ray absorptiometry actually measures the area density of bones \cite{lab6}, which is primary work for further information extraction. Area density is usually expressed in $g/c{m^2}$, it is used to describe the bone mass per unit of projected area, or the average mass per pixel. In order to diminish the error introduced by the soft tissues, dual-energy method is proposed in the measurement \cite{lab22,lab23}. For each projection, two images at different levels of energy are obtained with the dual-energy method.
For human body, the penetration of dual-energy X-ray can be expressed by
 \begin{eqnarray}
\label{eq2}
{I_L} = {I_{0L}}\exp \left[ { - {{(\mu } \mathord{\left/
 {\vphantom {{(\mu } \rho }} \right.
 \kern-\nulldelimiterspace} \rho }{)_{LS}}{M_S} - {{(\mu } \mathord{\left/
 {\vphantom {{(\mu } \rho }} \right.
 \kern-\nulldelimiterspace} \rho }{)_{LB}}{M_B}} \right].
\end{eqnarray}
 \begin{eqnarray}
\label{eq2}
{I_H} = {I_{0H}}\exp \left[ { - {{(\mu } \mathord{\left/
 {\vphantom {{(\mu } \rho }} \right.
 \kern-\nulldelimiterspace} \rho }{)_{HS}}{M_S} - {{(\mu } \mathord{\left/
 {\vphantom {{(\mu } \rho }} \right.
 \kern-\nulldelimiterspace} \rho }{)_{HB}}{M_B}} \right].
\end{eqnarray}
where subscript $L$ and $H$ represent low and high energy, respectively; and subscript $B$ and $S$ represent bone and soft tissue, respectively. Supposing that
${{\rm{U}}_{LS}}{\rm{ = }}{\left( {{\mu  \mathord{\left/
 {\vphantom {\mu  \rho }} \right.
 \kern-\nulldelimiterspace} \rho }} \right)_{LS}}$,
 ${U_{HS}} = {\left( {{\mu  \mathord{\left/
 {\vphantom {\mu  \rho }} \right.
 \kern-\nulldelimiterspace} \rho }} \right)_{HS}}$,
 ${U_{LB}} = {\left( {{\mu  \mathord{\left/
 {\vphantom {\mu  \rho }} \right.
 \kern-\nulldelimiterspace} \rho }} \right)_{LB}}$,
 ${U_{HB}} = {\left( {{\mu  \mathord{\left/
 {\vphantom {\mu  \rho }} \right.
 \kern-\nulldelimiterspace} \rho }} \right)_{HB}}$,
 ${R_S} = {U_{LS}}/{U_{HS}}$,
 ${R_B} = {U_{LB}}/{U_{HS}}$,
 the area density of the soft tissue and the bone can be obtained  from Eq. (4) and (5) as follows
\begin{eqnarray}
 \label{eq2}
 {M_S} = \frac{{{R_B}\ln ({I_H}/{I_{0H}}) - \ln ({I_L}/{I_{0L}})}}{{{U_{LS}} - {U_{HS}}/{R_B}}}.
\end{eqnarray}
\begin{eqnarray}
 \label{eq2}
{M_B} = \frac{{{R_S}\ln ({I_H}/{I_{0H}}) - \ln ({I_L}/{I_{0L}})}}{{{U_{LB}} - {U_{HB}}/{R_S}}}.
\end{eqnarray}

Accordingly, we are capable of making derivation of the formulas to calculate the area densities with dual-energy refraction data. The phase shift of X-ray when passing through an object is
\begin{eqnarray}
 \label{eq2}
\Phi  =  - k\int {\delta dt}.
\end{eqnarray}
where $k = {{2\pi } \mathord{\left/
 {\vphantom {{2\pi } \lambda }} \right.
 \kern-\nulldelimiterspace} \lambda }$ is the wave vector, $\delta $ is the phase factor and $t$ is the transmission length.
 If the phase object is homogeneous, namely, the phase factor $\delta $ is evenly distributed, Eq. (8) can be rewritten as
\begin{eqnarray}
 \label{eq2}
 \Phi  =  - k\delta t.
\end{eqnarray}
Likewise, we define mass phase factor as $\delta {\rm{/}}\rho $. Since we know that the area density $M = \rho t$, Eq. (9) is further transformed to
\begin{eqnarray}
 \label{eq2}
\Phi  =  - k(\delta {\rm{/}}\rho )M.
\end{eqnarray}
As described before, the sample cylinders are even along length direction. Consequently, the phase shift $\Phi $ is simply the function of the radius direction (we assume it as $x$ direction), the relationship between phase shift $\Phi $ and refraction angle $\alpha $ is correspondingly given by
\begin{eqnarray}
 \label{eq2}
\Phi  = k\int {\alpha (x)dx} .
\end{eqnarray}
Combining with the definition and assumption as is stated before, the area density function of the refraction angle is
\begin{eqnarray}
 \label{eq2}
M =  - \frac{{\left[ {\int {\alpha (x)dx} } \right]}}{{\delta {\rm{/}}\rho }}.
\end{eqnarray}
Similarly, the penetration of dual-energy X-ray through human body can be described by
\begin{eqnarray}
 \label{eq2}
{\Phi _L} =  - k{(\delta /\rho )_{SL}}{M_S} - k{(\delta /\rho )_{BL}}{M_B}.
\end{eqnarray}
\begin{eqnarray}
 \label{eq2}
{\Phi _H} =  - k{(\delta /\rho )_{SH}}{M_S} - k{(\delta /\rho )_{BH}}{M_B}.
\end{eqnarray}
Supposing that
${\Delta _{SL}} = {(\delta /\rho )_{SL}}$, ${\Delta _{SH}} = {(\delta /\rho )_{SH}}$, ${\Delta _{BL}} = {(\delta /\rho )_{BL}}$, ${\Delta _{BH}} = {(\delta /\rho )_{BH}}$, ${Z_S} = \frac{{{\Delta _{SH}}}}{{{\Delta _{SL}}}}$, ${Z_B} = \frac{{{\Delta _{BH}}}}{{{\Delta _{BL}}}}$,
then the area density of the soft tissue and the bone eventually become
\begin{eqnarray}
 \label{eq2}
{M_B} = \frac{{{\Phi _L}{Z_S} - {\Phi _H}}}{{k{\Delta _{BL}}({Z_B} - {Z_S})}}.
\end{eqnarray}
\begin{eqnarray}
 \label{eq2}
{M_S} = \frac{{{\Phi _L}{Z_B} - {\Phi _H}}}{{k{\Delta _{SL}}({Z_S} - {Z_B})}}.
\end{eqnarray}

\section{Materials and methods}

\subsection{Experimental setup and working principles}

For the experimental setup, a Talbot-Lau interferometer is used in combination with a tungsten rotating anode X-ray tube and a CCD detector constructed at the Institute of Multidisciplinary Research for Advanced Materials in the Tohoku University, Japan as illustrated by Fig. 1.
\begin{center}
\includegraphics[width=8cm]{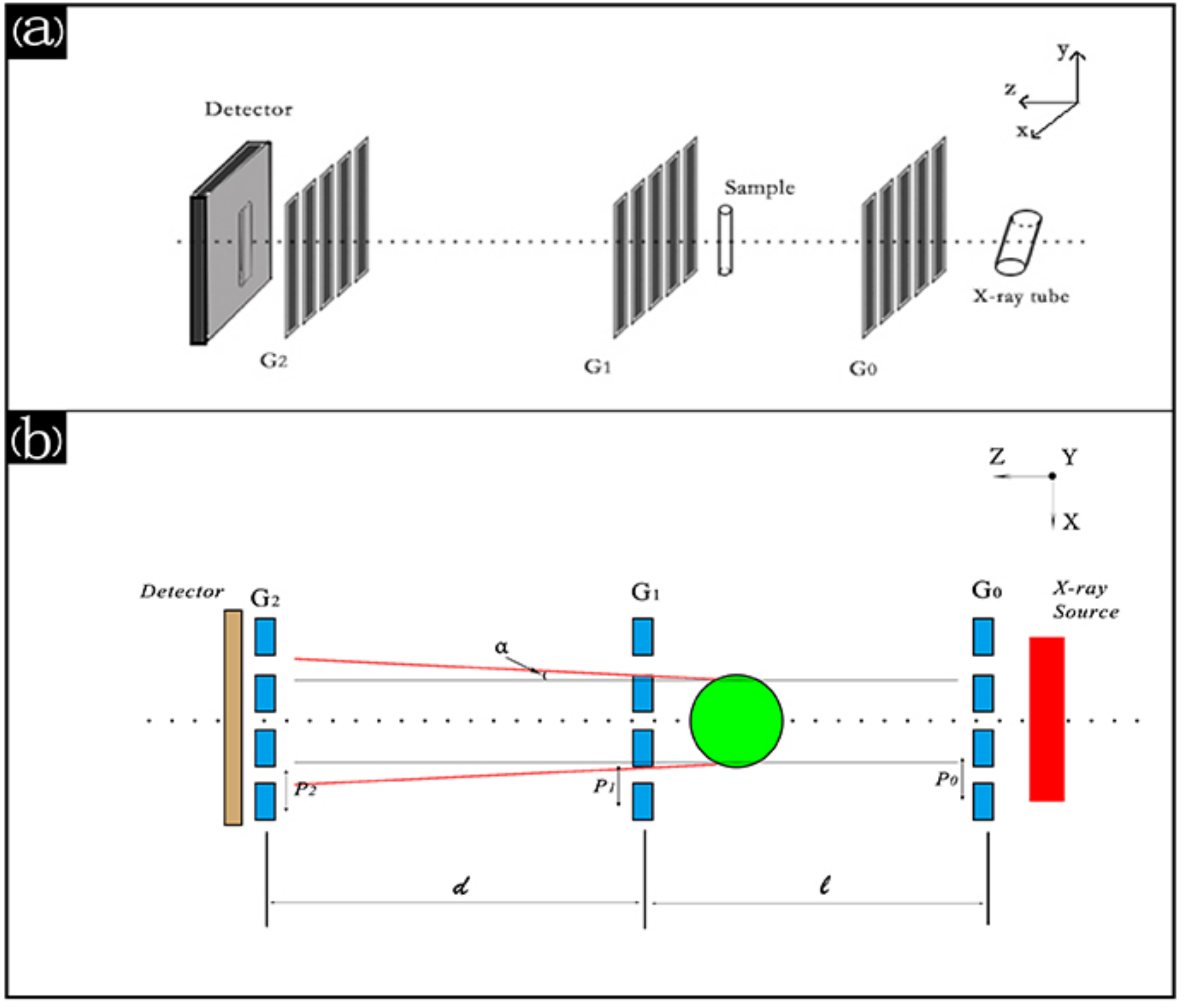}
\figcaption{\label{fig1}   (color online) Layout of hard-X-ray Talbot-Lau interferometer. (a), The scheme of Talbot-Lau interferometer. (b), The working principle of Talbot-Lau interferometer. }
\end{center}
The working current of the X-ray tube is 45 $mA$ , the source grating ${G_0}$ (period ${P_0} = 22.7$ $\mu m$, gold line height ${\rm{70}}$ $\mu m$, size ${\rm{20}} \times {\rm{20}}$ $m{m^{\rm{2}}}$) is made of gold, about 80 $mm$ downstream of the X-ray tube. The phase grating ${G_1}$ (period ${P_1} = 4.36$ $\mu m$, gold line height ${\rm{2}}{\rm{.43}}$ $\mu m$, size ${\rm{50}} \times {\rm{50}}$ $m{m^{\rm{2}}}$) is situated 106.9 $mm$($l$) downstream of the source grating. The phase grating is designed to generate a ${\pi  \mathord{\left/
 {\vphantom {\pi  {\rm{2}}}} \right.
 \kern-\nulldelimiterspace} {\rm{2}}}$ phase shift at an photon energy of 27 $keV$. The samples are positioned upstream of the phase grating with a distance of about 50 $mm$. The distance between the analyzer grating ${G_2}$ (period ${P_2} = 5.4$ $\mu m$, gold line height ${\rm{65}}$ $\mu m$, size ${\rm{50}} \times {\rm{50}}$ $m{m^{\rm{2}}}$) and the phase grating ${G_1}$ is 25.6 $mm$($d$). The detector, a scintillator CCD camera with an effective receiving size of ${\rm{68}}{\rm{.4}} \times {\rm{68}}{\rm{.4}}$ $m{m^2}$ and pixel size of ${\rm{18}} \times {\rm{18}}$ $\mu {m^2}$, is placed downstream of the analyzer grating ${G_2}$.

It is necessary to emphasize that the parameters of the optical elements are precisely designed and their relative positions are well optimized so as to obtain images of high quality. The working principles are referred to \cite{lab21}. The refraction angle $\alpha$ in the direction of axis $x$ can be quantified by \cite{lab24}
\begin{eqnarray}
\label{eq2}
\alpha = \frac{\lambda }{{\rm{2}}\pi }
\frac{{\partial \Phi (x,y)}}{{\partial x}}.
\end{eqnarray}
where $x$ and $y$ are the coordinates of the plane perpendicular to the optic axis, $\lambda $ is the wavelength of the incident X-ray and $\Phi (x,y)$ represents the phase shift of the wavefront.

\subsection{Data acquisition}

Three phase objects are used for quantitative measurement and evaluation. One PMMA rod is 5 $mm$ in diameter and two POM rods are in 5 $mm$ and 10 $mm$ in diameter, respectively. Three cylinders are all 100 $mm$ in height. During the experimental operation, all of them are oriented parallel to the grating lines. 5-step phase stepping scan is adopted in the experiment. At first, the accelerating voltage of the tube is set at 35 kV, and the samples are placed at the optical axis. Then the samples are removed, background images at 35 kV are obtained. Thereafter, the accelerating voltage ramps up to 45 kV, and the procedure of data acquisition is exactly the same.

\subsection{Information extraction}

The absorption signal and the refraction signal  are retrieved by the following equations \cite{lab25}:
\begin{eqnarray}
\label{eq2}
A(m,n) = \ln\left[\frac{{\sum\limits_{k = 1}^N {I_k^s(m,n)}}}{{\sum\limits_{k = 1}^N {I_k^b(m,n)}}}\right].
\end{eqnarray}
\begin{eqnarray}
\label{eq2}
\alpha (m,n) = \frac{{{P_2}}}{{2\pi d}} \times \arg \left[ {\frac{{\sum\limits_{k = 1}^N {I_k^s(m,n) \times \exp (2\pi i\frac{k}{N})} }}{{\sum\limits_{k = 1}^N {I_k^b(m,n) \times \exp (2\pi i\frac{k}{N})} }}} \right].
\end{eqnarray}
where $(m,n)$ denotes the pixel position of the CCD sensor, $N$ is the number of the steps, subscript $k$ represents that the data is obtained at ${k^{th}}$ step, superscript s and b denotes that the samples are added and removed , respectively. ${P_{\rm{2}}}$ is the period of the analyzer grating ${G_2}$, and $d$ is the distance between ${G_1}$ and ${G_2}$.

\section{Results}

\subsection{Information extraction}

Fig. 2 shows the absorption images and refraction images obtained from Eq. (18) and (19).
\begin{center}
\includegraphics[width=8cm]{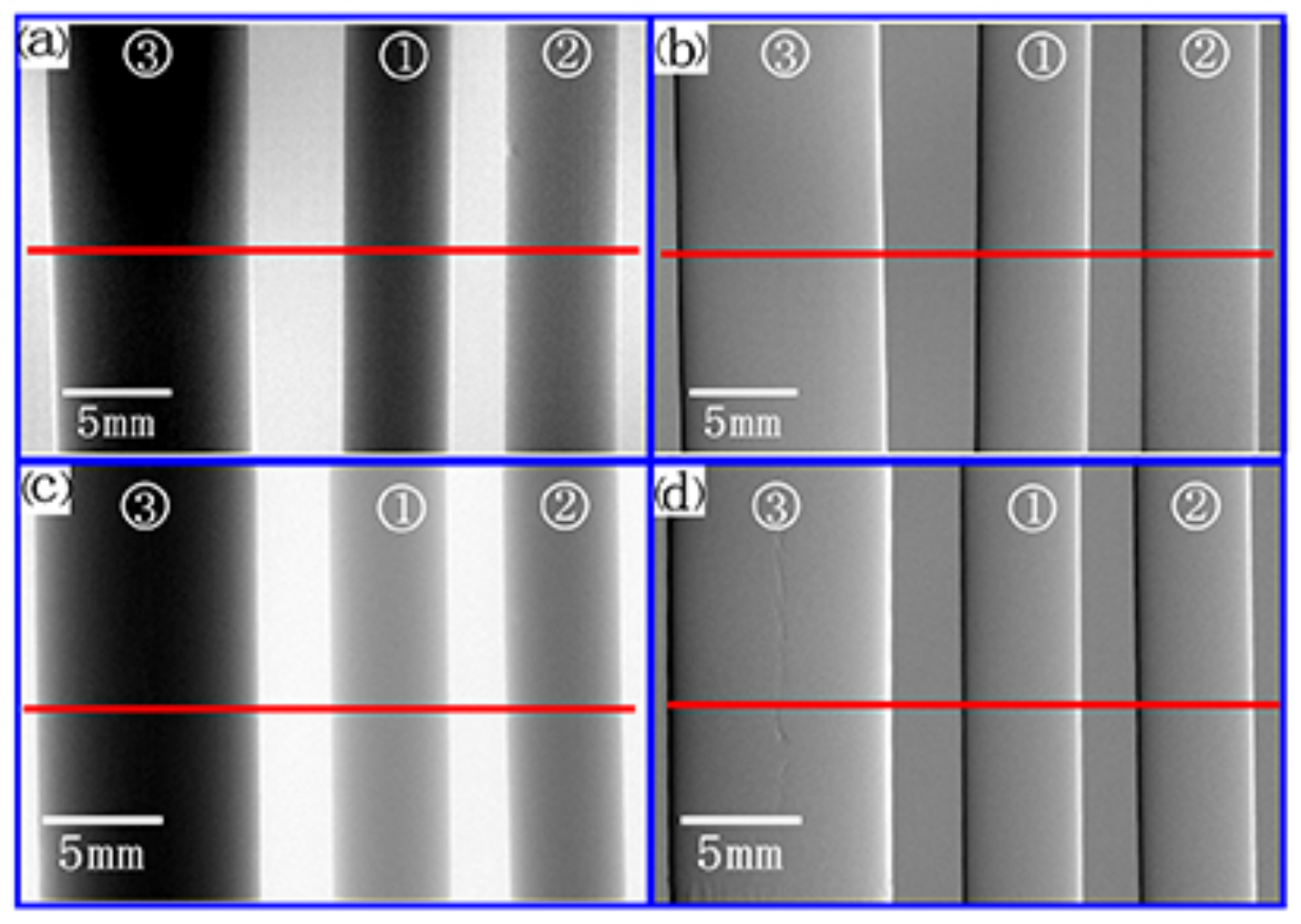}
\figcaption{\label{fig2}   (color online) Absorption and refraction images of the samples. (a), Absorption image at the tube voltage of 35 kV. (b), Refraction image at the tube voltage of 35 kV. (c), Absorption image at the tube voltage of 45 kV. (d), Refraction image at the tube voltage of 45 kV. The rods are also numbered, where ① is for the PMMA rod, ② is for the POM rod with the diameter of 5 mm and ③ is for the POM rod with the diameter of 10 mm). All images are windowed for optimized appearance with a linear gray scale.}
\end{center}

Specially, Fig. 2(a) shows the absorption image at the accelerating voltage of 35 kV, while Fig. 2(b) is its counterpart of the refraction image; Fig. 2(c) shows the absorption image at the accelerating voltage of 45 kV, and Fig. 2(d) is its counterpart of the refraction image.

\subsection{Single-energy calculation}

At first, we can calculate the area density of the samples with single-energy phase-contrast method. A cross section of the samples is selected for the calculation, which is denoted in Fig. 2 with red lines. The results are plotted in Fig. 3, and the parameters necessary for the calculation are listed in Table 1.

\end{multicols}
\ruleup
\begin{center}
\includegraphics[width=15cm]{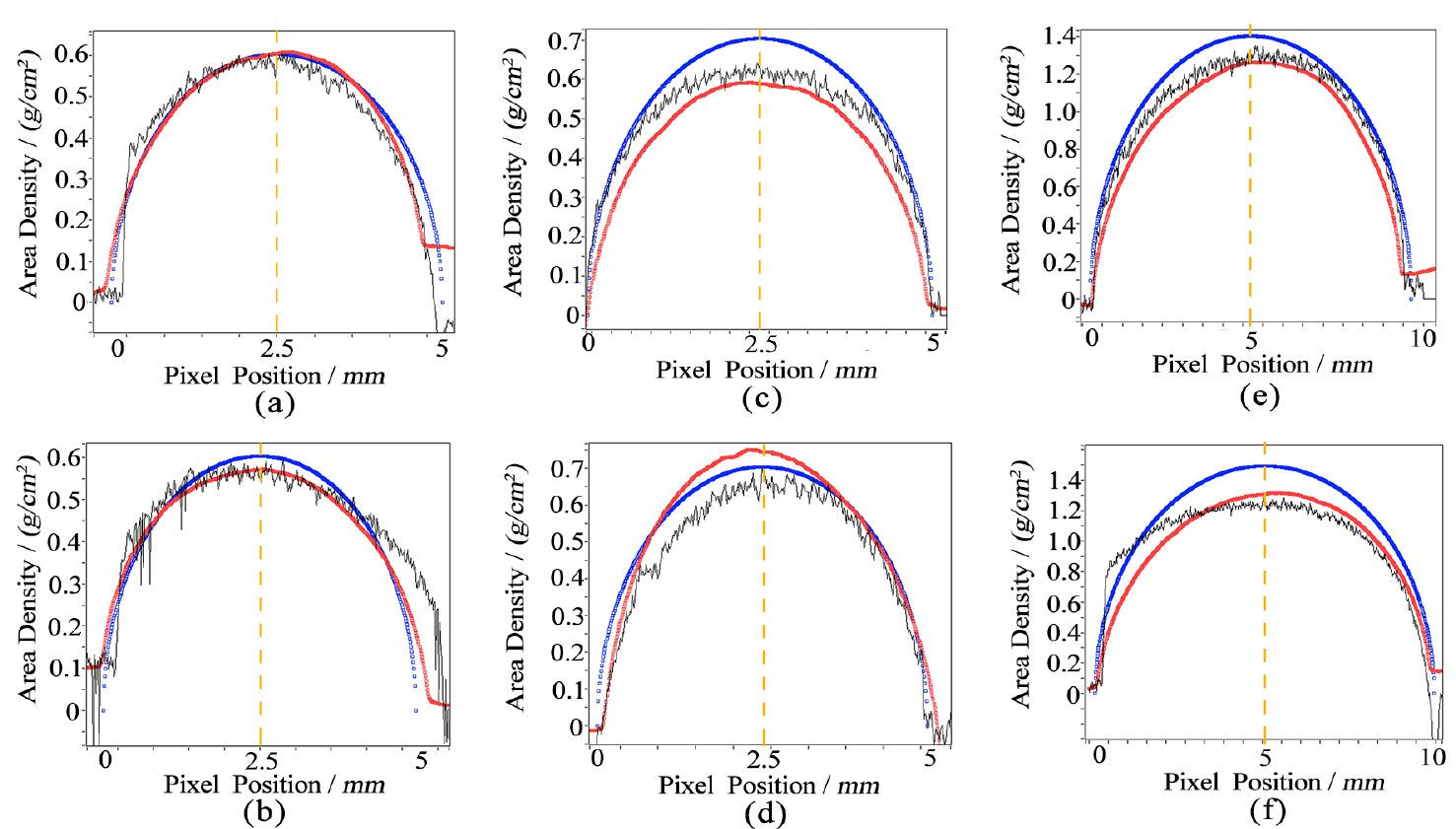}
\figcaption{\label{fig3}   (color online) The results for the area density calculation along the cross section with absorption method, refraction method and theoretical results, which is shown with black lines, red lines and blue lines respectively. (a), The results of sample ① at the tube voltage of 35 kV. (b), The results of sample ① at the tube voltage of 45 kV. (c), The results of sample ② at the tube voltage of 35 kV. (d), The results of sample ② at the tube voltage of 45 kV. (e), The results of sample ③ at the tube voltage of 35 kV. (f), The results of sample ③ at the tube voltage of 45 kV.}
\end{center}
\ruledown

\begin{multicols}{2}

\end{multicols}

\begin{center}
\tabcaption{ \label{tab1}  The absorption and refraction factor at different energies and the density of the samples \cite{lab26}.}
\footnotesize
\begin{tabular}{@{}ccccc@{}}\toprule
\multirow{2}{*}{Tube Voltage(Mean Energy)}&
\multicolumn{2}{c}{PMMA}  &
\multicolumn{2}{c}{POM}      \\
\cline{2-5}
&Absorption $\beta $($ \times {\rm{1}}{{\rm{0}}^{{\rm{ - 10}}}}$)&Refraction $\delta $($ \times {\rm{1}}{{\rm{0}}^{{\rm{ - 7}}}}$)& Absorption$\beta $($ \times {\rm{1}}{{\rm{0}}^{{\rm{ - 10}}}}$)& Refraction$\delta $($ \times {\rm{1}}{{\rm{0}}^{{\rm{ - 7}}}}$)                            \\
\midrule
35 kV(25 keV)          &1.53018681      &4.26404711       &2.03051909        &5.0097696     \\
45 kV(27.5 keV)        &1.22495458      &3.52414105       &1.60430406        &4.1531670     \\
\hline
$\rho/(g/c{m^3}) $         &
\multicolumn{2}{c}{1.19}  &
\multicolumn{2}{c}{1.42}      \\
\bottomrule
\end{tabular}
\end{center}

\begin{multicols}{2}

The black and red curves in Fig. 3 plot the area densities calculated based on conventional absorption method and the proposed phase-contrast method, respectively, in comparison with the theoretical prediction highlighted by the blue curve.
In order to quantify the calculation errors, the centric positions of the samples are chosen for comparison, which is also denoted in Fig. 3 with yellow dash lines. The values and mathematical errors are listed in Table 2. In the table, superscript $Theo$ and $Cal$ represent theoretic values and calculation values, respectively, ${E_L}$ and ${E_H}$ represents the low energy (the tube voltage 35 kV) and the high energy (the tube voltage 45 kV), respectively.

\end{multicols}

\begin{center}
\tabcaption{ \label{tab1}  The area density comparison between the calculation values and the theoretic values at centric positions of the samples.}
\footnotesize
\begin{tabular}{@{}lccccccccc@{}}\toprule
\multirow{3}{*}{Samples}&
\multirow{3}{*}{${M^{Theo}}/(g/c{m^2})$}&
\multicolumn{4}{c}{Absorption Method}  &
\multicolumn{4}{c}{Refraction Method}      \\
\cline{3-10}
 & &
\multicolumn{2}{c}{${E_L}$}  &
\multicolumn{2}{c}{${E_H}$}  &
\multicolumn{2}{c}{${E_L}$}  &
\multicolumn{2}{c}{${E_H}$}                 \\
\cline{3-10}
 & &
 ${M^{Cal}}/(g/c{m^2})$ &\%Error&${M^{Cal}}/(g/c{m^2})$ &\%Error&${M^{Cal}}/(g/c{m^2})$ &\%Error&${M^{Cal}}/(g/c{m^2})$ &\%Error    \\
\midrule
①PMMA  &0.595  &0.58 &-2.52  &0.57 &-4.20 &0.60 &0.84 &0.56 &-5.88      \\
②POM   &0.71   &0.62 &-12.6  &0.69 &-2.82 &0.57 &-19.7 &0.74 &4.23       \\
③POM   &1.42   &1.3 &-8.45  &1.19 &-16.1 &1.25 &-11.9 &1.32 &-7.04       \\
\bottomrule
\end{tabular}
\end{center}

\begin{multicols}{2}

\subsection{Dual-energy calculation}

This subsection investigates the feasibility of the dual-energy phase-contrast method described by Eq. (15) and (16) aiming to compute the area densities of two materials simultaneously. A numerical simulation is carried out on a contrast phantom whose cross section is shown in Fig. 4.
\begin{center}
\includegraphics[width=8cm]{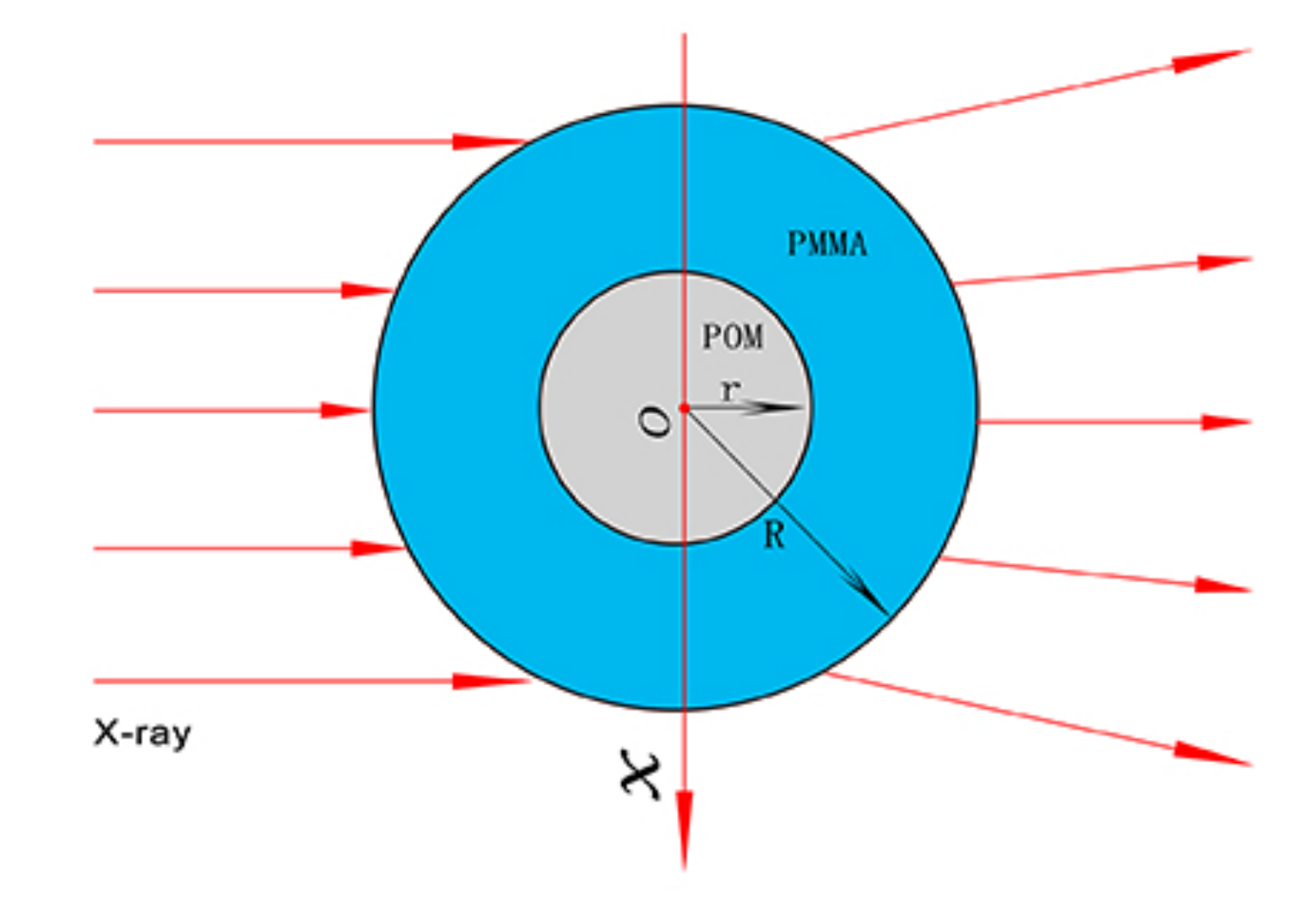}
\figcaption{\label{fig4}   (color online) The cross section of the contrast phantom used for the simulation. }
\end{center}

As the illustration shows, the phantom consists of a core cylinder made of POM, closely encircled by a cylindrical chamber made of PMMA ($r = 5$ $cm$, $R = 10$ $cm$). As before, we still set the tube voltage at 35 kV and 45 kV. Thus, the refraction angle of the X-ray $\theta (x)$ can be described as \cite{lab23}
\begin{eqnarray}
 \label{eq2}
\theta (x){\rm{ = }}\left\{ \begin{array}{l}
\displaystyle - \frac{{2{\delta _{PMMA}}x}}{{\sqrt {{R^2} - {x^2}} }},r \le \left| x \right| < R\\
~\\
\displaystyle \frac{{2{\delta _{PMMA}}x}}{{\sqrt {{r^2} - {x^2}} }} - \frac{{2{\delta _{PMMA}}x}}{{\sqrt {{R^2} - {x^2}} }} - \frac{{2{\delta _{POM}}x}}{{\sqrt {{r^2} - {x^2}} }},\left| x \right| < r
\end{array} \right.
\end{eqnarray}
where we indicate the radius direction as axis $x$ ,  the center of the cross section as origin as shown in Fig. 4.
To simplify the simulation, we define
\begin{eqnarray}
 \label{eq2}
\Theta {\rm{ = }}\int {\theta (x)dx}
\end{eqnarray}

Fig. 5 is the schematic representation of the simulation process. Fig. 5(a) shows the variation of the refraction angle according to the pixel position at the tube voltage of 35 kV and Fig. 5(b) is the integration thereof. Fig. 5(c) and 5(d) is the corresponding simulation results at the tube voltage of 45 kV.
\begin{center}
\includegraphics[width=8cm]{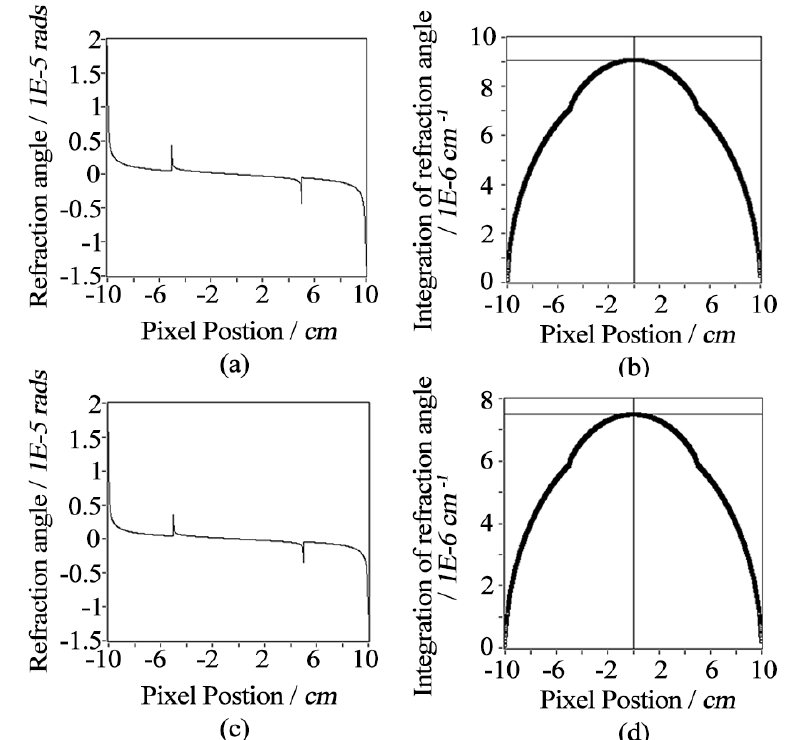}
\figcaption{\label{fig5}   Sketch of the simulation process. (a), Simulation of refraction angle at the tube voltage of 35 kV. (b), Simulation of the integration of refraction angle at the tube voltage of 35 kV. (c), Simulation of refraction angle at the tube voltage of 45 kV. (d), Simulation of the integration of refraction angle at the tube voltage of 45 kV.}
\end{center}

Again the centric position of the phantom is selected for comparison with the theoretic values. As Fig. 5 shows,
${\Theta _L} = 9.08765 \times {10^{ - 6}}$, ${\Theta _H} = {\rm{7}}{\rm{.50131}} \times {10^{ - 6}}$.
Based on the data in Table 1, we can figure out that
${\Delta _{SL}}{\rm{ = }}{(\delta /\rho )_{SL}}{\rm{ = 3}}{\rm{.58323287}} \times {\rm{1}}{{\rm{0}}^{{\rm{ - 7}}}}$,
${\Delta _{SH}}{\rm{ = }}{(\delta /\rho )_{SH}} = 2.96146307 \times {10^{ - 7}}$,
${\Delta _{BL}}{\rm{ = }}{(\delta /\rho )_{BL}} = 3.60390141 \times {10^{ - 7}}$,
${\Delta _{BH}}{\rm{ = }}{(\delta /\rho )_{BH}} = 2.97178169 \times {10^{ - 7}}$,
${Z_S} = \frac{{{\Delta _{SH}}}}{{{\Delta _{SL}}}} = 0.82647798$,
${Z_B} = \frac{{{\Delta _{BH}}}}{{{\Delta _{BL}}}} = 0.82460127$,
consequently,

~\\
${M_{POM}} =  \displaystyle\frac{{{\Theta _L}{Z_S} - {\Theta _H}}}{{{\Delta _{BL}}({Z_B} - {Z_S})}} = 13.94({g \mathord{\left/
 {\vphantom {g {c{m^{ - 2}}}}} \right.
 \kern-\nulldelimiterspace} {c{m^{ - 2}}}})$
 ~\\
${M_{PMMA}} =  \displaystyle\frac{{{\Theta _L}{Z_B} - {\Theta _H}}}{{{\Delta _{SL}}({Z_S} - {Z_B})}} = 11.335({g \mathord{\left/
 {\vphantom {g {c{m^{ - 2}}}}} \right.
 \kern-\nulldelimiterspace} {c{m^{ - 2}}}})$
~\\

According to our assumption, however, the theoretical values ought to be

~\\
${M_{POM}} = {\rho _{POM}} \times {D_{POM}} = 14.2({g \mathord{\left/
 {\vphantom {g {c{m^{ - 2}}}}} \right.
 \kern-\nulldelimiterspace} {c{m^{ - 2}}}})$
 ~\\
${M_{PMMA}} = {\rho _{PMMA}} \times {D_{PMMA}} = 11.9({g \mathord{\left/
 {\vphantom {g {c{m^{ - 2}}}}} \right.
 \kern-\nulldelimiterspace} {c{m^{ - 2}}}})$
 ~\\

where ${D_{POM}}$ and ${D_{PMMA}}$ represents the transmission length of the core cylinder and the chamber cylinder at the centric position respectively.
Naturally, the simulation errors are calculated as follows.

~\\
${\Delta _{POM}} =  \displaystyle\frac{{M_{POM}^{Cal} - M_{POM}^{Theo}}}{{M_{POM}^{Theo}}} \times 100\%  =  - 1.83\% $
~\\
${\Delta _{PMMA}} =  \displaystyle\frac{{M_{PMMA}^{Cal} - M_{PMMA}^{Theo}}}{{M_{PMMA}^{Theo}}} \times 100\%  =  - 4.75\% $
~\\

where superscript $Cal$ represents simulation values and superscript $Theo$ represents the theoretic values.

\section{Discussion}

Fig. 4 shows the calculation results with the absorption method and the novel phase-contrast method whereas Table 2 quantifies the corresponding calculation errors, which both proves the validity of the proposed method (calculation error is $<$5\% generally). Moreover, the errors can result from: (1) the parameters gauge of the apparatus, the optical elements and the samples; (2) the absorption and refraction of air. Furthermore, it is worth noticing that there is a fearful noise jamming in the absorption data, conversely, the refraction data follow a relatively smooth curve. The distinction between the calculation processes is able to explain the divergence. In the refraction calculation method, an integration is contained, which is actually a filtering operation, however the absorption method is based on the original images, making the absorption method inferior to the refraction method in noise decrease. Nevertheless, calculation errors can also be introduced by the integration, further works is to quantify the errors.

As stated before, the new dual-energy method is devoted to the accurate calculation of low-Z materials, so two low-Z materials close to each other in effective atomic number (${Z_{PMMA}} = 6.47$, ${Z_{POM}} = 6.95$) are employed in the numerical simulation. The simulation results is well consistent with theoretic values (simulation errors are $- 4.75\% $ and $- 1.83\% $, respectively), and it demonstrates that the new method is convincing. Likewise, the error sources can be analyzed: (1) theoretically, the refraction angle at the cylinder edge approaches infinity, so the simulation involves an approximation; (2) the integration to determine the phase shift involves errors.

X-ray phase-contrast imaging possesses evidently better contrast compared to the conventional absorption-based imaging for low-Z materials, meanwhile they perform similarly for high-Z materials. So the dual-energy phase-contrast method has certain advantages over the dual-energy absorption method, and they can briefly summarized as: (1) lower the radiation dose. With the novel dual-energy phase-contrast method, we can utilize X-ray source with lower power and lower photon energy to obtain desirable images and calculation accuracy. Meanwhile, the radiation dose is reduced, which is critical if we aim to extend its utilization to clinical application. (2) improve the calculation accuracy for bones as well as the soft tissues. In bone density measurement, soft tissue density is also measured, and the measure accuracy directly affects that of the bones. With the dual-energy phase-contrast method, soft tissue density can be measured with an improved accuracy, and it thus improves the bone density measurement as well. (3) reduce or eliminate the utilization of contrast agent in certain operations. In the existing dual-energy subtraction digital radiography, contrast agent is requisite because of the similarity for soft tissues (e.g. the vessels and its surrounding tissues). The dual-energy phase-contrast method has an inherent merit in distinction of similar low-Z materials, so the contrast agent is not demanded anymore and the operation procedure is simplified.

However, constrained by the experimental conditions, only two approaching tube voltages are selected (35 kV and 45 kV). Therefore, the influence of the tube voltage on the calculation precision is not fully investigated and understood. Besides, the phase-contrast imaging method we employed is based on Talbot-Lau interferometer, and it is generally optimized for a certain photon energy. As a result, the alternation of the tube voltage decreases the image quality. Hence we hope to develop a flexible dual-energy phase-contrast imaging method in the future.

\section{Conclusion}

In this paper, a novel phase-contrast method for area density calculation is presented, including single-energy method and dual-energy method. Experimental results from the refraction images and a numerical simulation demonstrate that the new method is applicable to area density measurement. Further extension of the work includes assessing the error with experiment on biological samples, quantifying the radiation dose, optimizing the dual X-ray energies and developing a flexible dual-energy phase-contrast imaging system. Moreover, another research endeavor is to apply dual-energy phase-contrast subtraction digital radiography in cardiovascular disease or inchoate cancer diagnosis, which may revolutionize the fields.

~\\
\acknowledgments{Authors would like to extend our sincere thanks to CHEN Fei-Fei at National University of Singapore, who checked through the paper with patience and gave instructive suggestions. And authors also acknowledge the financial funding support from the Japan-Asia Youth Exchange program in Science (SAKURA Exchange Program in Science) administered by the Japan Science and Technology Agency.}

\end{multicols}

\vspace{-1mm}
\centerline{\rule{80mm}{0.1pt}}
\vspace{2mm}

\begin{multicols}{2}

\end{multicols}

\clearpage


\begin{thebibliography}{90}

\vspace{3mm}

\bibitem{lab1} G.N.Hounsfield. British Journal of Radiology, 1973,  {\bf 46}:  1016-1022

\bibitem{lab2} Alvarez R E and Macovski A. Physics in medicine and biology, 1976,  {\bf 21}:  733

\bibitem{lab3} Zatz L M. Radiology, 1976,  {\bf 119}:  683-688

\bibitem{lab4} Macovski A, Alvarez R, Chan J-H, et al. Computers in biology and medicine, 1976,  {\bf 6}:  325-336

\bibitem{lab5} R.A.Rutherford B R P, and Isherwood. Neuroradiology. 1976,  {\bf 11}:  15-21

\bibitem{lab6} Mazess R B, Barden H S, Bisek J P, et al. Am J Clin Nutr, 1990,  {\bf 51}:  1106-1112

\bibitem{lab7} Ole Lander Svendsen J H, Christian Hassager, and Claus Christiansen. The American journal of clinical nutrition, 1993,  {\bf 57}:  608-608

\bibitem{lab8} P.J.Ryan. Seminars in Nuclear Medicine, 1997,  {\bf 3}:  197-209

\bibitem{lab9} Glovanni Di Chiro M D, Rodney A. Brooks, Ph.D., Robert M. Kessier. work in progress, 1979,  {\bf 131}:  521-523

\bibitem{lab10} ZHANG L J, WU S Y, Poon C S, et al. Journal of computer assisted tomography, 2010,  {\bf 34}:  816-824

\bibitem{lab11} Macdonald R D R. Machine Vision Applications In Industrial Inspection Ix, 2001,  {\bf 4301}:  31-41

\bibitem{lab12} Ying Z R, Naidu R and Crawford C R. Journal Of X-Ray Science And Technology, 2006,  {\bf 14}:  235-256

\bibitem{lab13} Momose A, Takeda T, Itai Y, et al. Nature medicine, 1996,  {\bf 473-475}

\bibitem{lab14} Kottler C, Revol V, Kaufmann R, et al. Journal of Applied Physics, 2010,  {\bf 108}:  114906

\bibitem{lab15} Momose A. Japanese Journal Of Applied Physics Part 1-Regular Papers Brief Communications \& Review Papers, 2005,  {\bf 44}:  6355-6367

\bibitem{lab16} David C, No?hammer B, Solak H H, et al. Applied Physics Letters, 2002,  {\bf 81}:  3287

\bibitem{lab17} Davis T, GAO D, Gureyev T, et al. Nature, 1995,  {\bf 373}:  595-598

\bibitem{lab18} Wilkins S, Gureyev T, Gao D, et al. Nature, 1996,  {\bf 384}:  335-338

\bibitem{lab19} Momose A. Nuclear Instruments and Methods in Physics Research Section A: Accelerators, Spectrometers, Detectors and Associated Equipment, 1995,  {\bf 352}:  622-628

\bibitem{lab20} Snigirev A, Snigireva I, Kohn V, et al. Review of Scientific Instruments, 1995,  {\bf 66}:  5486-5492

\bibitem{lab21} Pfeiffer F, Weitkamp T, Bunk O, et al. Nature Physics, 2006,  {\bf 2}:  258-261

\bibitem{lab22} Ostlere S J and Gold R H. Clinical orthopaedics and related research, 1991,  {\bf 271}:  149-163

\bibitem{lab23} Dunn W, Wahner H and Riggs B. Radiology, 1980,  {\bf 136}:  485-487

\bibitem{lab24} Born M \& Wolf. Principles of Optics. Oxford: Pergamon, 1980

\bibitem{lab25} WANG S H, Margie P. Olbinado, Momose A et al. Chin. Phys. B, 2015 {\bf 24}:  68703-068703

\bibitem{lab26} Henke B L, Gullikson E M and Davis J C. Atomic data and nuclear data tables, 1993,  {\bf 54}:  181-342

\end{thebibliography}
\end{document}